\documentclass[12pt]{article}

\usepackage{latexsym}
\newif\ifams\amsfalse
\amstrue \ifams
 \usepackage{amsfonts}
\fi
\newcommand{\bea}{\begin{eqnarray}}
\newcommand{\eea}{\end{eqnarray}}
\newcommand{\be}{\begin{equation}}
\newcommand{\ee}{\end{equation}}

\ifams
  \else
 
\fi

\begin{document}
\title{
\begin{flushright}
\begin{small}
 Alberta-Thy 07-00\\
 May 2000 \\
\end{small}
\end{flushright}
\vspace{1.cm}
Holographic Stress Tensor for Kerr-AdS Black Holes and Local Failure on
IR-UV Connection}
\author{Jeongwon Ho\\
\small \vspace{- 3mm} Theoretical Physics Institute, Department of Physics,\\
\small \vspace{- 3mm} University of Alberta, Edmonton, Canada T6G 2J1\\
\small E-mail: {\tt jwho@phys.ualberta.ca}}
\date{}
\maketitle

\begin{abstract}
We show that in general holographic stress tensor may contain a new term
of divergence of a spacelike unit normal acceleration. Then, it is shown
that in contrast to previous descriptions, a new stress tensor for
Kerr-AdS solutions can be a traceless one. Interestingly, this prescription
entails a local failure on the IR-UV connection.
\end{abstract}
\bigskip
\newpage 

A precise form of the AdS/CFT correspondence \cite{mal} has been formulated
by Gubser, Polyakov, Klebanov \cite{gubser} and Witten \cite{witten}.
The basic statement is given by the correspondence
between the partition function of a string (or M) theory and the generating
functional of correlation functions of a boundary conformal field theory
(CFT). In the description, the two are functionals of boundary fields
$\phi^{(0)}$ which have dual perspective as boundary values of bulk fields
$\phi$ and sources for the operators of CFT. According to this view,
in the limit of large number of D-branes and small string coupling,
the effective action of strong coupling large $N$ CFT is given by evaluating
the action functional for solutions to classical supergravity equations of
motion
\be
\label{coreq}
S_{string}[\phi^{cl}(\phi^{(0)})] = W_{CFT}[\phi^{(0)}].
\ee

Concentrating on the stress tensor in CFT and the corresponding bulk gauge
field, metric $g_{\mu\nu}$, it may be assumed that all other bulk fields vanish
on the boundary. For AdS spaces, due to existence of a second order pole
on the asymptotic boundary, the metric $g_{\mu\nu}$ does not induce a unique
metric $g^{(0)}$ on the boundary. Instead, the boundary field that satisfies
the boundary condition of the bulk metric is a conformal structure
$[g^{(0)}]$. However, the conformal invariance is to be broken on the
regularization of the bulk supergravity action as arbitrarily picking
a particular representative $g^{(0)}$ of the conformal structure $[g^{(0)}]$.
Taking this scheme, Henningson and Skenderis \cite{henn} have shown that the
conformal anomaly in CFT (ultraviolet (UV) effect) arises from infrared (IR)
divergences in the bulk theory. This is an explicit example of the
IR-UV connection \cite{susskind2}, which applies to holographic theories
\cite{thooft}\cite{susskind}, and becomes a non-trivial check for the
conjectured AdS/CFT correspondence. Generalizations and applications of the
investigation has been studied in
\cite{odintsov}\cite{sken}\cite{imbimbo}\cite{robinson}\cite{haro}.

According to the equation (\ref{coreq}), divergences arising from the
supergravity action is the usual UV divergences in CFT. Thus, the
regularization for the supergravity action can be achieved by introducing
local counterterms. Compared to the reference background subtraction
method \cite{brown}\cite{horowitz}, the prescription, so call counterterm
subtraction method, is a nice way to regularize a gravitational action
apparently preserving general covariance. The counterterm subtraction
method has been developed in its own interest and applications
\cite{bal}\cite{hyun}\cite{emparan}\cite{kra}\cite{solo}\cite{nojiri}.

On the other hand, various black holes have been studied in the context of
the AdS/CFT correspondence, and some interesting observations have been made,
e.g., on electrically charged Reissner-Nordstr\"om-AdS \cite{cham},
rotating Kerr-AdS \cite{hawking}\cite{berman}\cite{landsteiner},
and Kerr-Newman-AdS black holes \cite{caldarelli}\cite{hawking2}.
In this paper, we are concerned about the IR-UV connection between
the Kerr-AdS black holes and boundary CFT's living on rotating Einstein
universes. This subject has been served by a manuscript \cite{awad} in
which the correspondence was probed by calculating the Casimir energies
and/or conformal anomalies from bulk theory using the counterterm
subtraction method.

An interesting observation in \cite{awad} has been
made for the five-dimensional Kerr-AdS and the dual ${\cal N}=4$ super
Yang-Mills (SYM) theory on a rotating Einstein universe in four dimensions.
In usual, when a conformal anomaly is present, the classical bulk action
contains a logarithmic divergence, which cannot be apparently canceled
by a counterterm. For the five-dimensional Kerr-AdS,
the stress tensor was not traceless, but logarithmic divergence did not
appear in the on-shell action. Nevertheless, the trace of stress tensor
precisely matched to that of the dual SYM. The authors argued that not only
the integrated conformal anomaly vanishes, but also the anomaly is
proportional to $\Box R$, where $R$ is the boundary scalar curvature.
Therefore, supplementing ordinary counterterm action with an additional
counterterm, then one obtains a new traceless stress tensor. They also
proposed that the different choices for counterterms corresponds to the
choice of different schemes for regularization in ordinary field theories
in four dimensions that is due to the freedom of taking an undetermined
coefficient of $\Box R$ in the anomaly.

However, this prescription for the paradox seems unreasonable. First of all,
in the one loop effective action of the ${\cal N}=4$ SYM on four-dimensional
rotating Einstein universe, the $\Box R$ term
in present has to be distinguished from the $\Box R$ term with an undetermined
coefficient, e.g., depending on the choices of minimally coupled and
conformally coupled scalars \cite{burgess}. The former is just the usual
logarithmic UV divergence in four dimensions $R^{ab}R_{ab}-R^2/3$.
The proportionality, $\Box R \propto R^{ab}R_{ab}-R^2/3$, is a special
property of the geometry of the four-dimensional rotating Einstein
universe\footnote{Our argument has been given under consideration of the
weak coupling calculation. This seems still available in the strong coupling,
because the free energy density at weak coupling is only different from that
for the strong coupling (and a high temperature limit) up to a constant
factor for a leading term \cite{gubser2}\cite{bal}\cite{kim}\cite{hawking2}.}.
Therefore, it appears that there is not a precise relationship between
addition of new counterterms and the choice of the undetermined coefficient
of $\Box R$ in the field theory.
Secondly, it has to be noted that the addition of new counterterms means
that there may be `pulsative counterterms' which could be turned on and off
depending on given boundary geometries and/or topologies. However, considering
the procedure of the derivation of counterterm action in \cite{kra},
it must be available for all kinds of asymptotic AdS spaces with boundaries
of arbitrary geometries and topologies as solutions to the Einstein's
equations (containing only the gravitational field without other bulk fields).
Thus, it seems hard to put the pulsative counterterms into the counterterm
action with a consistent description. It has to do that just by hand.

In this paper, we revisit this paradox. Our starting point is to elaborate on
the on-shell action in the context of the ADM formulation. Taking into
account this description, we show that in general the stress tensor
may contain a new term of divergence of a spacelike unit normal
acceleration and be a traceless one. Then we shall argue that
this prescription interestingly may entail a local failure on the IR-UV
connection; One loop effective action of the ${\cal N}=4$ SYM on
four-dimensional rotating Einstein universe is UV finite, and correspondingly
the effective action evaluated from bulk action is IR finite. However,
the modified stress tensor derived from bulk theory may be traceless, while
the SYM has non-vanishing trace of stress tensor.

$(d+1)$-dimensional gravitational action with cosmological
constant $ \Lambda =- d(d-1)/(2\ell^2)$ is given by
\begin{equation}
\label{bact}
S = \frac{1}{16 \pi G} \int_{X} d^{d+1} x \sqrt{-g}
\left( \hat{R} + \frac{d(d-1)}{\ell^2} \right)
-\frac{1}{8 \pi G} \int_{\partial X}d^d x \sqrt{-\gamma} \Theta ,
\end{equation}
where $\partial X$ denotes $d$-dimensional boundary manifold with
metric $\gamma_{ab}$ and $\Theta_{ab}$ is the extrinsic curvature of
the boundary defined by $\Theta_{ab} = - \gamma_a^\mu \nabla_\mu n_b $.
$\nabla $ denotes the covariant derivative on the $(d+1)$-dimensional manifold
$X$ and $n^\mu $ is an outward unit normal to the boundary.
$\hat{R}$ is the bulk scalar curvature. The surface term in (\ref{bact}),
so called Gibbons-Hawking term, is required for well defined variational
principle. In this paper, we consider the bulk metric of the form
\begin{equation}
\label{metric}
g_{\mu\nu}dx^{\mu} dx^{\nu} = N^2dr^2 + \gamma_{ab}dx^a dx^b,
\end{equation}
where $x^r$ is the radial coordinate $r$, and $N^2 =N^2(r,x^a)$,
$\gamma=\gamma(r, x^a)$. In this coordinate system, the spacelike unit
normal to the boundary is given by $n_\mu = N \delta^r_\mu $.

According to the counterterm subtraction method, we introduce a counterterm
action ${\tilde S}$ regularizing the action (\ref{bact})
\bea
\label{adsct}
{\tilde S} &=& -\frac{1}{8\pi G}\int_{\partial X}d^dx\sqrt{-\gamma}
\left\{
\frac{d-1}{\ell} + \frac{\ell}{2(d-2)}R \right.
\nonumber \\
&& \left. + \frac{\ell^3}{2(d-2)^2(d-4)}\left(
R_{ab}R^{ab} - \frac{d}{4(d-1)}R^2 \right) +
 \cdots \right\}.
\eea
Then, the regularized action $S_p$ is defined by $S_p \equiv S + {\tilde S}$.
The line element of Kerr-AdS solutions ($d \geq 3 $) interested in this
paper is \cite{hawking}
\bea
\label{kerrbh}
ds^2 &=&
-\frac{\Delta_r}{\rho^2}
       \left(dt - \frac{a\sin^2{\theta} d\phi}{\zeta}\right)^2
+\frac{\Delta_\theta \sin^2{\theta}}{\rho^2}
       \left(adt - \frac{(r^2 + a^2)}{\zeta}d\phi \right)^2
\nonumber \\
&& +\frac{\rho^2}{\Delta_r}dr^2 + \frac{\rho^2}{\Delta_\theta} d\theta^2
+ r^2 \cos^2{\theta} d\Omega_{d-3}^2,
\eea
where
\bea
\label{metcoe}
\rho^2 &=& r^2 + a^2 \cos^2{\theta},~~
\Delta_r = (r^2+a^2)(1+r^2/\ell^2) - 2mGr^{4-d},
\nonumber \\
\zeta &=& 1 - a^2/\ell^2,~~
\Delta_\theta = 1 - (a^2/\ell^2)\cos^2{\theta},
\eea
and $m$ and $a$ denote the black hole mass and angular momentum per
unit mass, respectively. This is an AdS version of higher dimensional
Kerr black holes \cite{myers}.

The on-shell regularized action $S_p^{cl}$ of the
five-dimensional Kerr-AdS in (\ref{kerrbh}) does not contain a logarithmic
divergence \cite{awad}. Divergent part of the on-shell action is given by
\bea
\label{onshell}
S^{cl}_{div} &=& - \frac{1}{8 \pi G} \int_{X} d^{d+1} x \sqrt{-g}
\frac{d}{\ell^2} -\frac{1}{8\pi G}\int_{\partial X}d^d x \sqrt{-\gamma}\Theta ,
\\
&=& \int_{\partial X} d^d x \frac{\sqrt{\Omega_{d-3}}}{8 \pi G}r^{d-2}
\left[ \frac{(d-1)}{\ell^2}r^2 + (d-1) \left( 1 + \frac{a^2}{\ell^2}
 \left( 1 - \frac{2\cos^2{\theta}}{d-2} \right) \right) \right.
\nonumber \\
&&\left. -\frac{a^4}{r^4}\cos^2{\theta}\sin^2{\theta} \Delta_{\theta}+
\cdots + \frac{a^2(-a^2\cos^2{\theta})^{(d-6)/2}}{r^{d-4}}
\sin^2{\theta} \Delta_\theta \right]
 \frac{\sin{\theta}\cos^{d-3}{\theta}}{\zeta}. \nonumber
\eea
In (\ref{onshell}) and hereafter, we set $m=0$ in the metric
(\ref{kerrbh}). The terms including the mass is in fact finite on the
asymptotic region and is irrelevant to aim of this paper. In addition,
it is a necessary condition for counterterms that must be given in terms of
only intrinsic boundary geometry.

On the other hand, the divergence structure of the on-shell action
is tightly constrained by the Gauss-Codazzi equations in the sense that
these are covariant expressions given in terms of the intrinsic and
extrinsic boundary geometry \cite{kra}. Thus, we expect that the ADM
formulation gives us a hint for resolving the paradox above mentioned.
In fact, using the ADM formulation, the on-shell action can be expressed
by only the intrinsic boundary geometry up to redefinition of the
radial coordinate \cite{ho}. In this sense, we calculate
the on-shell action again in the context of the ADM formulation.

The canonical form of the action (\ref{bact}) is
\be
\label{canoact}
S =\frac{1}{16\pi G} \int_{X} d^{d+1} x
N\sqrt{-\gamma} \left(\Theta^2 - \Theta^{ab}\Theta_{ab}+
R + \frac{d(d-1)}{\ell^2} \right).
\ee
On the other hand, the Einstein equations contracted by the bulk metric
can be written by
\be
\label{aaeq}
   \Theta^2 - \Theta^{ab}\Theta_{ab} - R - \frac{d(d-1)}{\ell^2} = 0,
\ee
and
\be
\label{bbeq}
\Theta^{ab}\Theta_{ab} - n^\mu \nabla_\mu \Theta
- \nabla_\mu b^\nu- \frac{d}{\ell^2}= 0,
\ee
where $b^\mu \equiv n^\nu \nabla_\nu n^\mu $ is an acceleration of the
unit normal $n^\mu$. The first equation (\ref{aaeq}) is the
normal-normal component of the Gauss-Codazzi equations \cite{birrell}
and the second (\ref{bbeq}) can be identified with the tangential-tangential
one (requiring the equation (\ref{aaeq})). Substituting (\ref{aaeq}) into
(\ref{canoact}), the on-shell action is given by
\be
\label{clact}
S^{cl}=\frac{1}{8\pi G} \int_{X} d^{d+1} x
N\sqrt{-\gamma} \left( R + \frac{d(d-1)}{\ell^2} \right).
\ee
It must be noted that since we are concerned about the divergence structure
of the on-shell action, the equation (\ref{bbeq}) is irrelevant in our
calculation \cite{kra}. However, the term of divergence of the acceleration
in (\ref{bbeq}) is to play an important role in our prescription.

Now, we find that divergent part of the on-shell action (\ref{clact})
for $d$-dimensional Kerr-AdS solutions ($d=4,6,\cdots$) contains a
logarithmic term
\bea
\label{clactkerr}
 S^{cl}_{div} &=& \int_{\partial X} d^d x
\frac{\sqrt{\Omega_{d-3}}}{8 \pi G}r^{d-2} \left[
\frac{(d-1)}{\ell^2}r^2 + (d-1) \left( 1 + \frac{a^2}{\ell^2}
 \left( 1 - \frac{2\cos^2{\theta}}{d-2} \right) \right) \right.
\nonumber \\
&&\left. + \frac{a^2}{r^2}\left(d-3
 + \frac{2((d-3)\sin^2{\theta} - \cos^2{\theta})}{d-4}
- \frac{2a^2\cos^2{\theta}((d-2)\sin^2{\theta}-\cos^2{\theta})}{\ell^2(d-4)}
\right)\right.
\nonumber \\
&&\left.-\frac{a^4}{r^4}\left(\frac{2\cos^2{\theta}((d-3)\sin^2{\theta}
- \cos^2{\theta})}{d-6} - \frac{2a^2\cos^4{\theta}
((d-2)\sin^2{\theta}-\cos^2{\theta})}{\ell^2(d-6)} \right)
\right. \nonumber \\
&& \left. + \cdots + 2(-\cos{\theta})^{(d-4)}\ln r
\left(\frac{a}{r}\right)^{(d-4)}\left((d-3)\sin^2{\theta} -\cos^2{\theta}
\right.\right. \nonumber \\
&& \left.\left.-\frac{a^2\cos^2{\theta}}{\ell^2}((d-2)\sin^2{\theta}-\cos^2{\theta})
\right)\right] \frac{\sin{\theta}\cos^{d-3}{\theta}}{\zeta}.
\eea
The logarithmic divergence term in (\ref{clactkerr}) apparently leads
a conformal anomaly
\bea
\label{ano}
{\cal A} &=& \frac{-a^2(-a^2 \cos^2{\theta})^{(d-4)/2}}{8 \pi G} \times
\nonumber \\
&&\left(\frac{
   (d-3)\sin^2{\theta} - \cos^2{\theta}
    - a^2\cos^2{\theta}((d-2)\sin^2{\theta} - \cos^2{\theta})/\ell^2}{
\rho r^{d-3}\Delta_{r}^{(m=0)}} \right),
\eea
where we used a cutoff $r^2/\ell^2$ (c.f. \cite{emparan}).
As expected, for the case of $d=4$
the conformal anomaly in the leading contribution, ${\cal A}_{d=4}$,
is equal to that evaluated in \cite{awad}
\be
\label{ano4}
{\cal A}_{d=4}= -\frac{a^2\ell}{8 \pi G r^4}\left(\frac{a^2\cos^2{\theta}}{
\ell^2}(3\cos^2{\theta} - 2) - \cos{2\theta} \right).
\ee

Finally, we are in position of describing the discrepancy of the
on-shell actions in (\ref{onshell}) and (\ref{clactkerr}). Following the above
observation, especially deriving the conformal anomaly (\ref{ano4}),
the discrepancy should be closely related to the paradox given in \cite{awad}
why the stress tensor is not traceless, while the on-shell action
(\ref{onshell}) does not contain a logarithmic divergence.

The origin of the discrepancy is found in the canonical form of the action
(\ref{canoact}). In fact, it contained two total derivative terms, one canceled
the Gibbons-Hawking term, and the other was discarded by a simple algebraic
relation given by
\bea
\label{bdt1}
S_{bt} &=& \frac{1}{8 \pi G}
\int_X d^{d+1}x \sqrt{-g}\nabla_\mu (n^\nu\nabla_\nu n^\mu)
\nonumber \\
 &=& \frac{1}{8 \pi G}
\int_{\partial X} d^d x \sqrt{-\gamma} n_\mu(n^\nu\nabla_\nu n^\mu) = 0.
\eea
However, it is easily shown that the above calculation is not correct.
In the coordinate system of (\ref{metric}), the divergence of
acceleration $\nabla_\mu b^\mu$ cannot be a surface term of the timelike
boundary $\partial X$,
because it is not given by a total derivative term of the
radial coordinate
\be
\label{acc}
\sqrt{-g}\nabla_\mu b^\mu = -\partial_a (\sqrt{-\gamma}\gamma^{ab}
\partial_b N) = -\sqrt{-\gamma}D^a D_a N,
\ee
where $D_a$ is the covariant derivative defined on the timelike boundary.
Thus, we have to keep this term in calculation of the on-shell action
(\ref{clact})
\be
\label{clact1}
S^{cl}=\frac{1}{8\pi G} \int_{X} d^{d+1} x
N\sqrt{-\gamma} \left( R + \frac{d(d-1)}{\ell^2} + \nabla_\mu b^\mu \right).
\ee
The on-shell action (\ref{clact1}) for Kerr-AdS solutions
does not contain the logarithmic divergence term and recover that in
(\ref{onshell}).
According to this observation, it appears that the stress tensor
is modified by the term of divergence of unit normal acceleration
and becomes a traceless one.

Definition of the stress tensor is
\be
\label{dfst}
T^{ab} \equiv  \frac{2}{\sqrt{-\gamma}}\frac{\delta S^{cl}}{
\delta \gamma_{ab}}
= \frac{1}{8 \pi G} \left(\Theta^{ab} - \gamma^{ab} \Theta \right).
\ee
Actually, in the second equality, it was assumed that $\gamma^{\mu\nu}\delta
n_\mu = \delta \gamma^{\mu\nu} n_\mu=0$. This means that the boundary is fixed
under the variations so that the variations of the normal dual-vector on the
boundary are proportional to the normal dual-vector. (For an example, see
\cite{creighton}.) However, this assumption is no longer proper for the type
of metric (\ref{metric}), e.g., Kerr-AdS solutions, in which the radial lapse
$N$ is a function of a boundary coordinate as well as the radial one,
and this restriction has to be relaxed. Unfortunately, taking the relaxation
of the assumption, direct calculation of the stress tensor seems not easy
because of the particular algebraic form of the divergence
of the acceleration. Taking into account the form of the on-shell action
\ref{clact1}, we propose a form of new stress tensor as
\bea
\label{newst}
T^{ab}_{new} &=& \frac{1}{8 \pi G} \left( \Theta^{ab} - \gamma^{ab} \Theta
+ \frac{\alpha}{2} \gamma^{ab} \int dr N \nabla_\mu b^\mu 
\right ) \nonumber \\
&=& \frac{1}{8 \pi G} \left( \Theta^{ab} - \gamma^{ab} \Theta
- \alpha \gamma^{ab}K \right ),
\eea
where $K \equiv\frac{1}{2} \int dr D^cD_cN$. (In the following, we are calling
the terms that are just intuitively related to the $K$ term as `$K$ term'.)
If the constant $\alpha $ in (\ref{newst}) is one, then the new regularized
stress tensor $T^{ab}_p \equiv T^{ab}_{new} + \tilde{T}^{ab}$ is
traceless.

The usual scheme, which an on-shell action does not contain a logarithmic
divergence then the trace of stress tensor vanishes, make us lead to take
$\alpha =1$. In fact, there is another reason why $\alpha =1$ is attractive.
Kerr black hole solutions, which are asymptotically flat spacetime, can be
obtained by taking the flat spacetime limit $\ell \rightarrow \infty$
in \ref{kerrbh}. In the case, one can see that the `old' definition of the
stress tensor (\ref{dfst}) has a non-vanishing trace in leading contribution,
which is the same with the anomaly in (\ref{ano}) taking the flat spacetime
limit \cite{ho}. (In the calculation, the counterterm action for $d=4$ is
the form of
\be
\label{ctactft}
\tilde{S} = - \frac{1}{8 \pi G} \int_{\partial X} d^4x \sqrt{-\gamma}
\sqrt{\frac{3}{2} R},
\ee
(For the $d=4$ case, the counterterm action
(\ref{ctactft}) is enough to eliminate the divergence appearing in the
classical action\footnote{For higher dimensional Kerr solutions, see
\cite{kra}\cite{solo}}.) and the counterterm stress tensor $\tilde{T}^{ab}$
is given by
\be
\label{ctstft}
\tilde{T}^{ab} = \frac{1}{8 \pi G}\left(
\Phi(R^{ab} - \gamma^{ab}R) + D^aD^b \Phi - D^cD_c \Phi \gamma^{ab}
\right),
\ee
where $\Phi \equiv \sqrt{3/(2R)}$.) Taking into account the holographic
principle and extrapolating duality in the AdS/CFT correspondence, we
expact that there might be a quantum field theory living on the boundary
that is dual to a (super)gravity on an asymptotically flat spacetime.
However, it should not be a conformal field theory. Then, the stress tensor
with non-vanishing trace would be problematic. In this sense, the choice
of traceless stress tensor seems to be reasonable. The new stress tensor
in (\ref{newst}) with $\alpha =1$ taking the asymptotically flat limit
$\ell \rightarrow \infty $ gives a traceless regularized stress tensor.

Even though the new traceless stress tensor with $\alpha =1$ (\ref{newst}) is
plausible with the fact that the on-shell action does not contain logarithmic
divergence, it seems to be problematic on the AdS/CFT correspondence.
As mentioned above, the one loop effective action of the ${\cal N}=4$ SYM
on four-dimensional rotating Einstein universe is UV finite, so the equation
(\ref{coreq}) is still satisfied. However, the SYM has non-vanishing
conformal anomaly. In this sense, it appears that {\it the IR-UV
connection is locally broken}. In fact, this local failure occurs in the case
of $\alpha \neq 0$. In order words, the contribution of the $K$ term to
the stress tensor gives rise to the local failure on the
IR-UV connection.\footnote{It must be noted that the $K$ term does not
contribute to the total energy of the bulk theory in the leading contribution,
and the Casimir energy derived from the bulk theory still matches to
that of the boundary dual CFT.}
The $K$ term in the leading contribution is propertional to squared angular
momentum, and apparently, the local failure is due to rotation of the
bulk spacetime. In order to discuss this problem more,
we consider some aspects of the $K$ term in canonical point of view.

In some sense, the $K$ term measures how much deviated the boundary geometry
is from a round sphere. This reflects that the contribution appears in
the tangential-tangential component of the
Gauss-Codazzi equations (\ref{bbeq}).
On the other hand, the canonical form of the action (\ref{canoact}) including
the $K$ term is written in terms of canonical variables
\be
\label{canoact1}
S =\int_{X} d^{d+1}x \left(\pi^{ab}\gamma^{\prime}_{ab} - N{\cal H}
- \frac{\sqrt{-\gamma}}{8 \pi G} D^a D_a N \right),
\ee
where $\pi^{ab}$ is the conjugate momenta defined by $\pi^{ab} \equiv
\delta S/\delta \gamma^{\prime}_{ab}$. The radial Hamiltonian density
${\cal H}$ is given by
\be
\label{ham}
{\cal H} = \frac{16 \pi G}{\sqrt{-\gamma}}
\left( \frac{\pi^2}{d-1} - \pi_{ab}\pi^{ab} \right)
- \frac{\sqrt{-\gamma}}{16 \pi G} \left( R + \frac{d(d-1)}{\ell^2} \right).
\ee
Now, how can we understand the $K$ term in the canonical action
(\ref{canoact1})? First of all, the canonical action (\ref{canoact1})
can be rewritten by
\be
\label{canoact2}
S =\int dr \left[ \int_{\partial X} d^d x \left(\pi^{ab}\gamma^{\prime}_{ab}
- N{\cal H}\right) - \frac{1}{8 \pi G} \int_{\partial \partial X} d^{d-1} x
\sqrt{|h|} u^a D_a N \right],
\ee
where $h$ is an induced metric of a $(d-1)$-dimensional boundary $\partial
\partial X$ and $u^a$ is a unit normal to the $\partial \partial X$.
Thus, the $K$ term becomes a surface term of the Hamiltonian $H$
\be
\label{totham}
H = \int_{\partial X} d^d x {\cal H} + \frac{1}{8 \pi G} \int_{\partial
\partial X} d^{d-1} x \sqrt{|h|} u^a D_a N.
\ee
In usual, a surface term of a Hamiltonian plays an important role
in physics, e.g., as the total energy of a system. In this paper,
we have not find a physical description for this surface term. According to
the AdS/CFT correspondence, the Hamiltonian constraint
${\cal H}=0$ (turning on bulk scalar fields) is equivalent to the
renormalization group flow (RG-flow) equation of the boundary CFT
\cite{deboer}. However, the surface term should not give any contribution
on the CFT deformation, and moreover, we have been concentrated only on
the asymptotic boundary in which bulk scalar fields vanish. Thus, even though
it will be find that the surface term in (\ref{totham}) give a kind of
(local) deformation of boundary dual CFT related to the local
failure on the IR-UV connection,
it should be different from the CFT deformations recently studied in the
holographic RG-flow (For review, see \cite{petrini} and references
therein). We leave the investigation of prescription for a possible
description for the surface term in (\ref{totham}) and its relationship with
the local failure on the IR-UV connection as a future work. 

\vspace{0.2in} {\bf Acknowledgments:} I thank A. Zelnikov,
Y. Gusev, and S. Corley for helpful discussions. This work was supported by 
National Science and Engineering Research Council of Canada.

\end{document}